\documentclass[aps,prc,twocolumn,letterpaper,floatfix,nofootinbib]{revtex4}
\usepackage{graphicx,amsmath,amssymb}
\usepackage[usenames,dvipsnames]{xcolor}
\usepackage[usenames,dvipsnames]{pstricks}
\usepackage{pst-grad}
\usepackage{url}

\begin{document}

\title{Are there really two different Bell's theorems?}
\author{Travis Norsen}
\affiliation{tnorsen@smith.edu}

\date{February 24, 2015}

\begin{abstract}
This is a polemical response to Howard Wiseman's recent paper, ``The
two Bell's theorems of John Bell''.  Wiseman argues that, in 1964,
Bell established a conflict between the quantum mechanical
predictions and the joint assumptions of determinism and (what is now
usually known as)
``parameter independence''.  Only later, in 1976, did Bell, according to
Wiseman, first establish a conflict between the quantum mechanical
predictions and locality alone  (in the specific form that Bell would
sometimes call ``local causality'').  Thus, according to Wiseman, the 
long-standing
disagreements about what, exactly, Bell's theorem does and does not
prove can be understood largely as miscommunications resulting from the
fact that there are really two quite distinct ``Bell's theorems''.  My
goal here is to lay out what Wiseman briefly describes as an
``alternate reading'' of Bell's 1964 paper, according to which
(quoting Wiseman here) ``the first paragraph of Bell's `Formulation'
section [should be seen] as an essential part of his 1964 theorem, the
first part of a two-part argument.''    I will argue in particular
that this ``alternate reading'' is the correct way to understand
Bell's 1964 paper and that Wiseman's reading is strongly
inconsistent with the available evidence. 
\end{abstract}

\maketitle

\section{Introduction}

My goal here will be to summarize and record my side of a debate that
has erupted in response to Howard Wiseman's recent paper on ``The two
Bell's theorems of John Bell.'' \cite{wiseman} The debate
concerns the question (much in the air as we celebrate the 50th
anniversary of Bell's 1964 theorem) of what, exactly, Bell did and
didn't prove in 1964. \cite{bell1964} My view, which seems to match that of Bell
himself as well as several other contributors to this forum
(including for example Tim Maudlin and Jean Bricmont), is that already in 1964
Bell demonstrated the need for non-locality in any theory able to
reproduce the standard quantum predictions.  (``Non-locality'' here
means a violation of a generalized prohibition on faster-than-light
causal influences.)  Whereas the opposing view (which is probably a
majority view among normal physicists who have not studied Bell's work
carefully, and is especially prominent among those physicists Wiseman
describes as ``operationalists'') is that Bell only established a
conflict between the empirical predictions of quantum theory and the
joint assumptions of locality and determinism.  

Those adopting the opposing view tend to retain allegiance to locality
(which, they suggest, is after all a requirement of relativity) and
insist that the upshot of Bell's work is that determinism 
must be abandoned.  That is, they regard Bell's theorem as
fundamentally a no-hidden-variables proof and hence a vindication of
some standard (orthodox/Copenhagen/operationalist) 
interpretation of quantum mechanics.  This is
in contrast to Bell's own view, according to which the conflict with
experiment cannot be blamed on determinism or any other departure from
orthodoxy, but instead establishes that, however strongly motivated it
might be by relativity theory, locality is false.  

Wiseman's view, expressed in his recent paper, is neither of the above.  Instead, his view
is that both sides are right -- because they are talking about
different things.  In particular, according to Wiseman, the theorem in
Bell's 1964 paper shows exactly what the operationalists claim:
reproducing the quantum mechanical predictions requires abandoning
\emph{either} locality \emph{or} deterministic hidden variables.
Whereas, again according to Wiseman, Bell would \emph{later} (in 1976)
prove a second theorem establishing a conflict between the quantum
mechanical predictions and a unitary notion of ``local causality''
that captures the prohibition on faster-than-light causal influences
for general (stochastic, i.e., not necessarily deterministic)
theories.  So (according to Wiseman) Bell, Maudlin, Bricmont, myself,
and others are correct to say that ``Bell's theorem proves
non-locality'' -- if by ``non-locality'' we mean a failure of Bell's
``local causality'' condition --  while simultaneously those taking the
opposing view are right to say that ``Bell's theorem proves only that
deterministic hidden-variable theories have to be non-local.''  We're
both right; we just mean different things by ``Bell's theorem''.  It
was simple miscommunication all along. 

But I simply don't 
think that's \emph{right}.  The truth, I think, is that the people
taking the opposing view have simply missed, or misunderstood, the
role of the EPR argument in Bell's 1964 paper.\footnote{Or 
  they have refused, on some kind of anti-realist
  philosophical grounds, to even entertain as meaningful the issues
  that Einstein \emph{et al.} -- and Bell -- raised.}
And Wiseman, in
constructing an interpretation of that paper according to which the
opposing view is correct (even if only in regard to that
paper) essentially just repeats this common misunderstanding.  

There is a lot going on in Wiseman's paper and hence a lot that a full
analysis of it would need to go into. Providing such an analysis is
not my goal here.
Instead I will focus exclusively on the  question of what,
exactly, Bell wrote and meant and did in his 1964 paper.  It will turn
out that much of what is under dispute here hinges on
exactly what Bell meant by the word ``locality'' and, in particular,
on whether his several comments about ``locality'' should be
understood as attempts to provide a generalized 
definition of this term (Wiseman's view), or instead (my view) merely 
as descriptions of a narrower implication of locality in the context
of the particular type of theory that Bell took to have been
previously shown, by Einstein \emph{et al.}, to be required by a more
generalized notion of locality.  Fittingly, the dispute also involves
a disagreement about how to understand Bell's intentions with regard
to his repeated citation of a certain passage from Einstein's
``Autobiographical Notes''.  \cite{einstein}

In the following section I review Wiseman's reading of the 1964 paper
and then present, in the subsequent section, an overview of my own
reading.  A final section then elaborates on the several problems I
see in Wiseman's interpretation and summarizes the issues as I see them.  There is a
lot of quoting from Bell, Wiseman, and
Einstein. 
So to make reading this essay as easy as possible I have color-coded the
quotations from these three sources:  Bell (1964) in
\textcolor{blue}{blue}, Wiseman (2014) in \textcolor{red}{red}, and
Einstein (1949) in \textcolor{OliveGreen}{green}.  Quotations from other
sources are cited in the usual way.  

\section{Wiseman's Reading of Bell's 1964 Paper}

Wiseman describes Bell's 1964 theorem as showing \textcolor{red}{``that there are
quantum predictions incompatible with any theory satisfying locality
and determinism''} and emphasizes that \textcolor{red}{``Bell's 1964 theorem suggests
that Bell experiments leave us with a choice:  accept that physical
phenomena violate determinism, or accept that they violate locality.''}
As Wiseman acknowledges, this reading puts him in the ``almost
universal'' category of misunderstanding that Bell himself would later
call attention to:
\begin{quote}
``My own first paper on this subject ... starts with a summary of the
EPR argument \emph{from locality to} deterministic hidden variables.
But the commentators have almost universally reported that it begins
with deterministic hidden variables.''  \cite{bell1981}
\end{quote}
Nevertheless, Wiseman insists that his reading of the 1964 paper
(contrary to Bell's own later description of what he had done there)
is correct.  How does he justify this reading?

Wiseman begins by quoting Bell's several statements involving
``locality'' in the 1964 paper.  The first relevant passage occurs
in  \textcolor{blue}{``1  Introduction''} (which Wiseman and I agree is really more like an abstract):
\begin{quote}
\textcolor{blue}{``The paradox of Einstein, Podolsky, and Rosen was advanced as an
argument that quantum mechanics could not be a complete theory but
should be supplemented by additional variables.  These additional
variables were to restore to the theory causality and locality$^{[*]}$.
In this note that idea will be formulated mathematically and shown to
be incompatible with the statistical predictions of quantum
mechanics.  It is the requirement of locality, or more precisely that
the result of a measurement on one system be unaffected by operations
on a distant system with which it has interacted in the past, that
creates the essential difficulty.''}
\end{quote}
Bell's footnote references the following excerpt from  Einstein's
``Autobiographical Notes'' in the 1949 Schilpp volume:  \textcolor{OliveGreen}{``But on one supposition we
should, in my opinion, absolutely hold fast:  the real factual
situation of the system $S_2$ is independent of what is done with the
system $S_1$, which is spatially separated from the former.''}

Wiseman proceeds to assume \textcolor{red}{``that the `real factual situation' of a
system is what is probed by measuring it''} and notes that \textcolor{red}{``the
notions of being `independent of what is done with' or `unaffected by
operations on' a system clearly refer to the action of an agent (say
Alice) on her system, and mean that Alice's action has no statistical
effect.''}  It is not completely clear what, exactly, Wiseman takes the relationship to be,
between Bell's own words here and the words he quotes from Einstein. (This will be discussed extensively later.)  But Wiseman
does seem to allow his interpretation of Einstein's words to influence
his interpretation of Bell's words, and he does acknowledge that
\textcolor{red}{``Bell's definition of locality follows from the supposition of
Einstein's which [Bell] quotes.''}  In any case, all of this leads
Wiseman to formalize Bell's  definition of locality as:
\begin{equation}
P_{\theta}(B | a,b,c,\lambda) = P_{\theta}(B | b,c,\lambda)
\end{equation}
which is the same condition that is usually referred to as
``parameter independence'' (PI) in the more recent Bell
literature.\footnote{The ``$\theta$'' subscripts in the formula refer
  to the particular candidate theory assigning the probabilities in question.
Note also that the formula remains somewhat vague
  until one specifies exactly what each symbol -- and in particular
  the notoriously controversial $\lambda$ -- is meant to capture.
  Wiseman is not explicit about this, but seems to follow Bell in understanding
  the $\lambda$  (together with $c$) as denoting a complete specification of 
  the physical state of the particle pair
  at some appropriate time prior to any measurement.  But this is also
  slightly puzzling since Wiseman also seems to think that 
  \textcolor{red}{``[f]or the operationalist, locality [i.e., PI] is a
    natural assumption...''}  In my opinion, any
    genuine ``operationalist'' would simply balk at a notion of
    locality that required (to use Bell's later characterization)
    comparing probabilities conditioned on ``a full specification of
    local beables in a [certain] space-time region''.  \cite{bell1990}
    Thus I think to some extent Wiseman conflates 
    PI with a (distinct, and genuinely operationally meaningful) ``no
    signaling'' condition.  (See Ref. \cite{belljarrett} for some
    further relevant discussion.)  And I think this conflation is based to some
    extent on forgetting the $\theta$ subscripts in Equation (1),
    i.e., thinking that the probabilities this formula relates can be
    understood as empirical frequencies.  (Notice for example
    Wiseman's use of the phrase ``statistical effect''.)  But, having
    noted it here, I will ignore this side issue in the remainder of this paper.}

Wiseman also quotes Bell's summary statement, from the Conclusion of
his 1964 paper:  
\begin{quote}
\textcolor{blue}{``In a theory in which parameters are added to quantum
mechanics to determine the results of individual measurements, without
changing the statistical predictions, there must be a mechanism
whereby the setting of one measuring device can influence the reading
of another instrument, however remote.''}
\end{quote}
Wiseman italicizes the phrase \textcolor{blue}{``the setting of one measuring
device can influence the reading of another instrument, however
remote''} and explains that he regards this as constituting a \emph{formulation} of
\textcolor{red}{``(the negation of) locality''}.  Wiseman writes:  
\begin{quote}
\textcolor{red}{``As the above quote
shows, Bell definitely means locality specifically as the absence of
any influence of the \emph{setting a} on the remote measurement
device.  This confirms my above reading of his definition of locality
in [equation (1) above].    In fact, this reading is confirmed in
two more places in the paper.''}
\end{quote}
The two other places cited by Wiseman include the first paragraph of 
Bell's \textcolor{blue}{``2 Formulation''} where Bell writes:
\begin{quote}
\textcolor{blue}{``Now we make the hypothesis$^{[*]}$, and it seems one at least worth
considering, that if the two measurements are made at places remote
from one another the orientation of one magnet does not influence the
result obtained with the other.''}
\end{quote}
(Note that Bell again here cites the same passage from Einstein's
``Autobiographical Notes'' that was quoted
earlier.)  The other passage that Wiseman regards as confirming his
interpretation of what Bell means by ``locality'' comes later in
Bell's section 2:
\begin{quote}
\textcolor{blue}{``The vital assumption$^{[*]}$ is that the result \emph{B} for particle 2 does not
depend on the setting \emph{a} of the magnet for particle 1, nor
\emph{A} on \emph{b}.''}
\end{quote}
(Note that Bell here cites Einstein for a third time.)

Having thus laid out his evidence for interpreting Bell as having
meant, by ``locality'' in 1964, our Equation (1) above, Wiseman turns
his attention to Bell's recapitulation, in \textcolor{blue}{``2 Formulation''}, of the
EPR argument.  Wiseman quotes Bell's one-sentence summary of the
argument
\begin{quote}
\textcolor{blue}{``Since we can predict in advance the result of measuring any chosen
component of $\vec{\sigma}_2$ by previously measuring the same
component of $\vec{\sigma}_1$, it follows that the result of any such
measurement must actually be predetermined.''}
\end{quote}
and then remarks:  
\begin{quote}
\textcolor{red}{``Here Bell has made a mistake.  His conclusion
(predetermined results) does not follow from his premises
(predictability, and [PI]).  This is simple to see from the
following counter-example.  Orthodox quantum mechanics (OQM) is a
theory in which the setting $a$ of one device does not statistically
influence the result $B$ obtained with the other:
\begin{equation}
P_{\theta}(B|a,b,c,\lambda) = P_{\theta}(B|b,c,\lambda) \nonumber
\end{equation}
Here, if $c$ were to correspond to preparation of a mixed quantum
state $\rho_c$, the variable $\lambda$ would allow for a pure-state
decomposition; in purely operational QM, only $c$ would appear.  But
in OQM it is of course not true that the results of spin measurements
are predetermined for a singlet state as Bell is considering.''}  
\end{quote}
Wiseman is here pointing
out that OQM (which respects PI and is hence
``local'' in the sense Wiseman attributes here to Bell) predicts the
usual kind of perfect correlations in the EPR setup, but fails to
attribute outcome-determining local hidden-variables to the individual
particles in the EPR pair.  Wiseman, that is, regards orthodox quantum
mechanics as a counter-example to the EPR argument -- or, at least,
the recapitulation of it that Wiseman interprets Bell as giving here.

And that is basically that.  Wiseman closes by asserting (presumably
on the grounds that he considers the argument as presented invalid) 
that \textcolor{red}{``Bell's
EPR paragraph forms no part of his 1964 theorem''} and remarking as
follows on his accusation that Bell made a mistake in thinking that
the EPR argument (``\emph{from locality to} deterministic hidden
variables'') was valid:
\begin{quote}
\textcolor{red}{``I would classify Bell's mistake in this paragraph as a peccadillo,
having no impact on the main result in his paper.  It would have been
an easy mistake for Bell to have made, if he had the idea that EPR had
already proven determinism from some sort of locality assumption, and
did not think hard about whether it was the same as the locality
assumption he was about to use in his own theorem.  Indeed the paper
could be made completely sound by replacing `it follows' in the above
(`Since we can predict...') quote by `the obvious explanation is', or
`EPR's premises imply'.  Although Bell believed that he was
reproducing EPR's argument, EPR's premises (which are never stated by
Bell) are \emph{not} equivalent to locality (as defined here by Bell),
and they \emph{do} justify the conclusion of pre-determined
outcomes...''}
\end{quote}
(For a proof that EPR's premises, 
including a notion of locality distinct from the one Wiseman
attributes here to Bell, we are sent to an Appendix in
Wiseman's paper.)  

\section{My Own Reading}

Let me here give an overview of my own reading of Bell's 1964 paper,
and then come back (in the next section) to explain exactly what I
find implausible about Wiseman's interpretation.  

I would begin with something Wiseman seems to barely notice:  the
title of Bell's paper.  This, I think, already makes it quite obvious
that Bell intends his novel result to be understood as being built
``On the Einstein-Podolsky-Rosen paradox''.   That is, I think, Bell
takes himself (quite correctly) to be adding a crucial second step to
what had been previously established by Einstein \emph{et al.}   This
foundational  role of the EPR argument in Bell's work is made quite
clear in the first section, \textcolor{blue}{``1  Introduction''},
which I quote here in full:
\begin{quote}
\textcolor{blue}{``The paradox of Einstein, Podolsky, and Rosen was advanced as an
argument that quantum mechanics could not be a complete theory but
should be supplemented by additional variables.  These additional
variables were to restore to the theory causality and locality$^{[*]}$.  In
this note that idea will be formulated mathematically and shown to be
incompatible with the statistical predictions of quantum mechanics.
It is the requirement of locality, or more precisely that the result
of a measurement on one system be unaffected by operations on a
distant system with which it has interacted in the past, that creates
the essential difficulty.  There have been attempts to show that even
without such a separability or locality requirement no `hidden
variable' interpretation of quantum mechanics is possible.  These
attempts have been examined elsewhere and found wanting.  Moreover, a
hidden variable interpretation of elementary quantum theory has been
explicitly constructed.  That particular interpretation has indeed a
grossly non-local structure.  This is characteristic, according to the
result to be proved here, of any such theory which reproduces exactly
the quantum mechanical predictions.''}
\end{quote}
The second part of the paragraph here tells us something about Bell's
motivation for undertaking the reported work (namely, he wanted to see
if \emph{any} deterministic completion of quantum theory would
\emph{have} to have the \textcolor{blue}{``grossly non-local structure''} displayed by
the de Broglie - Bohm pilot-wave theory).  

But let us focus here on the first part, which (like Wiseman) I read as essentially
an abstract of the paper.  It begins by noting that, according to the
earlier EPR argument, a hidden-variable type theory could \textcolor{blue}{``restore
to the theory [i.e., quantum mechanics] causality and locality''.}  
This clearly implies that, according to Bell, Einstein \emph{et al.}
had previously established that ordinary quantum mechanics
\emph{violates} both \textcolor{blue}{``causality and locality''}.
The violation of causality (which, like Wiseman, I understand here to
simply mean ``determinism'') is uncontroversial and unremarkable.  But
it is important to appreciate that already here Bell is claiming
(and/or endorsing Einstein's previous claim) that ordinary quantum
mechanics violates ``locality''.   
This is certainly consistent with what we know of
Einstein's criticisms of quantum mechanics (to be elaborated further
in the following section).  In particular, it is consistent with the
passage from Einstein that Bell specifically chose to cite here (and
then two subsequent times) as,
evidently, capturing his (Bell's) own understanding of this concept:
\begin{quote}
\textcolor{OliveGreen}{``But on one supposition we should, in my opinion, absolutely hold
fast:  the real factual situation of the system $S_2$ is independent
of what is done with the system $S_1$ which is spatially separated
from the former.'' } 
\end{quote}
In my opinion, what Bell means by ``locality'' has thus \emph{already} been
made rather clear:  he means some sense of ``locality'' that (i) is
reasonably well-captured by the Einstein passage from
``Autobiographical Notes'', (ii) had been involved in the EPR
argument, and (iii) is violated by ordinary quantum mechanics.  
(Parameter Independence, of course, satisfies none of these three
criteria.)

But what about the immediately-following sentences
of Bell's \textcolor{blue}{``1  Introduction''}?   Here Bell writes:  \textcolor{blue}{``In this note that idea
will be formulated mathematically and shown to be incompatible with
the statistical predictions of quantum mechanics.  It is the
requirement of locality, or more precisely that the result of a
measurement on one system be unaffected by operations on a distant
system with which it has interacted in the past, that creates the
essential difficulty.''}   First off, what is \textcolor{blue}{``that idea''} which will
be formualted mathematically?  I read Bell here as referring, with
\textcolor{blue}{``that idea''}, back to the \emph{conjunction} of \textcolor{blue}{``causality and
locality''} -- i.e., the two features that were to be restored by the
introduction of additional variables.  This is, after all,
precisely what he does later formulate mathematically in the first
equation appearing in his paper.  
According to this equation, which I reproduce here, the outcomes are mathematically determined by
locally-accessible variables:
\begin{equation}
A(a,\lambda) = \pm 1 , \; B(b,\lambda) = \pm 1.
\end{equation}
I thus interpret the statement about locality from the abstract 
 (namely, \textcolor{blue}{``the result of a measurement on
  one system [should] be unaffected by operations on a distant system
  with which it has interacted in the past''}) not as an attempt to
give a general formulation or definition of locality (he has already
done this by quoting Einstein!), but instead
as a description of the specific implication of locality (to
deterministic hidden variable theories) that  he will use later in the body
of his paper. This seems perfectly natural since the
statement appears in (what amounts to) an abstract of the paper, i.e.,
in a summary of the novel result the paper will announce.   Note in particular that the statement in question
comes just after the future-tensed statement about what
\textcolor{blue}{``will be formulated mathematically''}.  Whereas the first
two sentences of the abstract refer to the earlier
work of EPR -- and Einstein's earlier formulation of locality -- in the past tense.

It is also natural to interpret, in this same way, Bell's statement
from later in the paper, just after he writes what I have transcribed
in equation (2) above:
\begin{quote}
\textcolor{blue}{``The vital assumption is that the result $B$ for particle 2 does not
depend on the setting $a$, of the magnet for particle 1, nor
$A$ on $b$.''} 
\end{quote}
Here, that is, he is not telling us what ``locality'' (in the most
general sense) means, but instead calling our attention to a
particular feature of the deterministic model he's just written down:
namely, it is a \emph{local} deterministic hidden-variable theory.
And similarly for his summarizing sentence in \textcolor{blue}{``6 Conclusion''}:
\begin{quote}
\textcolor{blue}{``In a theory in which parameters are added to quantum mechanics to
determine the results of individual measurements, without changing the
statistical predictions, there must be a mechanism whereby the setting
of one measuring device can influence the reading of another
instrument, however remote.''}
\end{quote}
Note in particular that the violation of locality (namely, the
existence of \textcolor{blue}{``a mechanism whereby the setting of one measuring device
can influence the reading of another instrument, however remote''}) is
described here as applying specifically to deterministic hidden
variable theories (i.e., theories \textcolor{blue}{``in which
  parameters are added to quantum mechanics to determine the results
  of individual measurements''}.)   

These three statements -- that Wiseman interprets as attempts to
\emph{define} ``locality'' -- thus instead seem to me to be clearly only
attempts to describe the specific implication of locality that Bell
uses in the context of the deterministic hidden-variable type theory
that, he argues (citing EPR), is required to restore locality to QM.

Thus, I think the overall structure of Bell's paper is as follows:
first he cites EPR as having previously established that locality (in
Einstein's sense) requires positing deterministic hidden variables (in
order to explain the predicted perfect correlations); then, in the
main body of the paper, he lays out his new proof that this kind
of local deterministic theory runs afoul of the quantum predictions
when more general correlations are considered.   It is crucial here
that Bell takes EPR to have \emph{previously established} the need for
deterministic hidden variables in order to restore locality.  Bell is
effectively (and, in retrospect, somewhat naively and unfortunately) 
taking for granted that his readers understand that this
has already been established, and is thus (quite reasonably) focusing
his expositional attention on the novel result that he has established, building on the
foundation laid by EPR.  If we drop this
overall context (i.e., ignore the foundational role of EPR and assume
that Bell is starting from scratch) we are likely to misinterpret much
of what he says, about ``locality'' in particular. 

There is, however, one statement about ``locality'' in Bell's paper
which is, from my point of view, somewhat problematic.   This occurs
in the first pargraph of \textcolor{blue}{``2 Formulation''} where Bell is
recapitulating the EPR argument (that he would later characterize as
an argument ``\emph{from locality to} deterministic hidden
variables''):
\begin{quote}
\textcolor{blue}{``With the example advocated by Bohm and Aharanov, the EPR argument is
the following.  Consider a pair of spin one-half particles formed
somehow in the singlet spin state and moving freely in opposite
directions.  Measurements can be made, say by Stern-Gerlach magnets,
on selected components of the spins $\sigma_1$ and
$\sigma_2$.  If measurement of the component $\sigma_1
\cdot a$, where $a$ is some unit vector, yields the
value $+1$ then, according to quantum mechanics, measurement of
$\sigma_2 \cdot a$ must yield the value $-1$ and vice
versa.  Now we make the hypothesis$^{[*]}$, and it seems one at least
worth considering, that if the two measurements are made at places
remote from one another the orientation of one magnet does not
influence the result obtained with the other.  Since we can predict in
advance the result of measuring any chosen component of
$\sigma_2$, by previously measuring the same component of
$\sigma_1$, it follows that the result of any such measurement
must actually be predetermined.''}  
\end{quote}
This is the only point in the paper where Bell is actually attempting
to recapitulate the logic of the EPR argument and hence explain
exactly why and how pre-determination really \textcolor{blue}{``follows''}
from locality and perfect correlations.  So it is here that we would
most want and expect to see an explicit \emph{general formulation} of
locality (rather than just some statement about one of locality's
implications in the specific context of deterministic theories).  And,
unfortunately, Bell disappoints us.  Other than citing Einstein, what he says here about locality
(\textcolor{blue}{``if the two measurements are made at places remote from one another
the orientation of one magnet does not influence the result obtained
with the other''}) certainly falls short of a general formulation
(along the lines that he would later give, in 1976 and 1990).  It
should be clear, for example, from the involvement of ``magnets'' that
he is only here talking about some kind of implication of locality in
the specific EPR-Bohm setup (with spin 1/2 particles whose spins are
measured using Stern-Gerlach magnets).  

But even leaving that
disappointing specificity aside, what Bell says here seems problematic
in another way as well:  what does it mean to say that some distant
intervention \textcolor{blue}{``does not influence the result obtained''} by a nearby
measurement?   As the following five decades of Bell literature
eloquently  illustrate, it is notoriously difficult and controversial
to precisely capture the
idea of causal influence in the context of general (not necessarily
deterministic) theories.\footnote{Indeed, it is not even really clear
  how one should relate ordinary quantum mechanics (with
  realistically-interpreted and collapsing wave functions) to what
  Bell writes here in words.  In ordinary QM, the orientation of the
  distant magnet certainly does influence the \textcolor{OliveGreen}{``real
    factual situation''} (on which much more later) of the nearby
  particle; and then the \textcolor{OliveGreen}{``real factual
    situation''} of that nearby particle certainly does
  \textcolor{blue}{``influence the result obtained''} in the nearby
  measurement.  But, because each of these influences involves some
  randomness, it turns out that the nearby measurement outcome is
  statistically independent of the distant setting (i.e., it turns out
  that PI is respected).  So, has the distant magnet setting
  influenced the nearby result?  It is simply not clear.}
So it is simply not clear how to translate Bell's
words here (about locality) into a sharp mathematical statement in
terms of which the EPR argument might be rigorously rehearsed.  

So, and especially taking into account the five decades of controversy
that have followed, it must be admitted that Bell's recapitulation of
the EPR argument in this paragraph leaves something to be desired.
And given the increasing attention that Bell gave to this very point
in his subsequent writings (for example, by later providing more fully
general mathematical formulations of the idea of ``locality'' and by
stressing more explicitly the precise arguments by which ordinary QM
can be seen to violate locality and by which deterministic hidden
variables can be seen to be genuinely required if the perfect correlations are to be
explained locally) it seems that Bell himself would agree that this
important aspect of his 1964 paper could and should have been
strengthened.  

But let us not lose sight of the big picture here.  The EPR argument
-- and the EPR-ish argument given by Einstein in the passages
surrounding the sentence cited \emph{three times} by Bell, including
in the very sentence we have just been scrutinizing -- were, in
the context of Bell's 1964 paper, ``prior work''.  Bell was (rightly
or wrongly) taking that prior work as given, taking its results as already established.  
And so even in the
important first paragraph of \textcolor{blue}{``2 Formulation''} we
should not understand him as attempting to present a fully rigorous
and detailed version of the argument 
(``\emph{from locality to} deterministic hidden variables'').
Instead, I think, we should understand him as giving a 
quick overview of this earlier argument, the fuller version of which
he invites his readers to find in the Einstein (\emph{et al.}) papers
(that is, the ``Autobiographical Notes'' and the EPR paper) which he
explicitly references.

In summary, I think that in 1964 Bell was taking for granted that
Einstein \emph{et al.} had previously established that determinism was
required in order to provide a local account of the perfect (EPR)
correlations.  The main new result Bell presented in 1964 was that
this particular method of attempting to restore locality to quantum
mechanics could not succeed, since local deterministic (hidden
variable) theories could not reproduce the QM predictions for a wider
class of possible experiments.  But to summarize the significance of
Bell's 1964 paper by saying that he demonstrated a conflict between
the QM predictions and the joint assumptions of ``locality'' and
``determinism'' is to simply ignore the crucial foundational role
played by the earlier work of Einstein \emph{et al}.  It is clear
that, for Bell, the significance of his new result was to show that
locality simply cannot be maintained if the quantum predictions are
correct:  \textcolor{blue}{``It is the requirement of locality
  ... that creates the essential difficulty.''}

\section{Discussion}

So, whose reading of Bell's 1964 paper is correct?  One important
piece of evidence in support of my reading is simply that it agrees
with what Bell himself later says about what he had been up to in
1964.  Of course, such testimony is only reliable to the extent that
there is independent evidence that Bell was an honest reporter about
his own earlier work.
But here there is literally universal agreement, among those who knew
him and worked with him, that Bell was an almost uniquely humble,
honest, and forthright person who took extreme care to get details
right and to always err on the side of crediting others rather than
himself.  I would also submit, as relevant evidence, Bell's 1977
remarks on ``Free variables and local causality'', which include the
following open confession of an earlier mistake (having
nothing directly to do with what's at issue here):  
``Here I must concede at once that the hypothesis becomes
quite inadequate when weakened in this way.  The theorem no longer
follows.  I was mistaken.''    \cite{bell1977}  Clearly Bell
had no difficulty admitting mistakes when he made them.  

Wiseman's interpretation, which requires one to believe that Bell made
a mistake in 1964 and then engaged in a decades-long terminological
cover-up campaign, simply does not seem plausible given what we know
about Bell.\footnote{Wiseman denies that he is
  \textcolor{red}{``accusing Bell or his followers of intellectual
    dishonesty''}.  But for me this is difficult to reconcile with
  his description of what must have happened subsequently (under
  the assumption, of course, that Wiseman is right about what Bell
  meant by ``locality'' in 1964):  \textcolor{red}{``once Bell had explicitly defined
    [local causality in 1976], he wished all previous localistic
    notions he had used, in particular the notion of [Parameter
    Independence], to be forgotten.  Moreover, after a few
    years he became convinced that it was the notion of [local}
   \textcolor{red}{ causality] that he had in mind all along.  [For example], Bell
    implies in 1981 that both he and Einstein were always using the
    notion of [local} \textcolor{red}{causality], which Bell characterises later in
    this 1981 paper in the same way [he had described it] in 1976.  As
    argued [previously] there is only one plausible reading of
    `locality' in Bell's 1964 paper, and it is not [local
    causality].''}  Needless to say, I find it very troubling that,
  for Wiseman, it is not even \emph{plausible} to consider that Bell
  might have meant, by ``locality'', the sort of condition given by
  Einstein in the passage he referenced, three times, apparently by
  way of telling us what he meant by ``locality''...  and that, by
  contrast, Wiseman does consider it entirely plausible (and indeed
  conclusively established) that Bell in effect lied, successfully, to
  himself about what he had meant.}
But there are many other and more direct reasons to
reject Wiseman's interpretation.

First and foremost, Wiseman's reading requires us to understand Bell
to have meant, by ``locality'', the condition that would later become
known as Parameter Independence.  This, I submit, is completely
and utterly implausible.  I previously noted that Wiseman's
interpretation fails to meet all three of the criteria that arise
already in the first two sentences of Bell's paper.  Furthermore, and
more even directly, nothing like my Equation (1) -- expressing PI -- 
appears anywhere in Bell's paper.  Nor does Bell say, in words,
anything that can in any direct sense be translated as PI -- which, of
course, is a statement about
\emph{probabilities}.  \emph{All} of the statements that Bell
makes (in his own voice) about locality in 1964 are statements about \textcolor{blue}{``the reading''} of
an instrument or \textcolor{blue}{``the result''} of an experiment.  That is, they are
statements that can only really be directly translated into
mathematics in the context of specifically deterministic theories.  I think that if
one really wanted to attempt to capture, in a mathematical expression,
what Bell says in words, it would look like this:
\begin{equation}
A(a,b,\lambda) = A(a,\lambda).
\end{equation}
In arriving instead at his mathematical translation, our Equation (1) above, of
Bell's various words, Wiseman is thus clearly engaging in some pretty creative
interpretation.  

Recall that Bell makes very clear, by citing Einstein three
different times, that his \emph{general} notion of locality -- as
opposed to the specific implication of it that he applies to
deterministic theories -- was the notion that Bell understood Einstein
to have in mind when he (Einstein) wrote:  
\begin{quote}
\textcolor{OliveGreen}{``But on one supposition we
should, in my opinion, absolutely hold fast:  the real factual
situation of the system $S_2$ is independent of what is done with the
system $S_1$, which is spatially separated from the former.''  }
\end{quote}
Does Wiseman thus mean to suggest that this passage also expresses
Parameter Independence?  
 As noted
earlier, Wiseman takes Einstein's \textcolor{OliveGreen}{``real
  factual situation''} of some system to denote
\textcolor{red}{``what is probed by measuring it''} and takes
Einstein's idea of a system being \textcolor{OliveGreen}{``independent''} of distant operations
as meaning that the distant \textcolor{red}{``action has no statistical
  effect''}. This strange, vaguely operationalist gloss on Einstein's
words seems suspicious to me, as if Wiseman is indeed trying to
suggest that Einstein, too, should be interpreted as having meant
Parameter Independence.  That, of course, would be ridiculous. 
But there is also an indication (when he
writes that \textcolor{red}{``Bell's definition of locality follows
  from the supposition of Einstein's which [Bell] quotes''}) that in
Wiseman's view
Einstein's notion of locality is distinct from and more generalized
than PI.  

But then why would Bell
specifically cite this \textcolor{red}{``supposition of Einstein's''}
\emph{three different times}, in contexts where it is clear that Bell
takes the passage to be explicating and clarifying the concept of
``locality'', if Bell actually meant, by ``locality'', something
distinct and narrower?  Wiseman simply never provides an answer to this
crucial question.

It is worthwhile to step back and look also at the
passages from Einstein's ``Autobiographical Notes'' that surround
the partial sentence quoted by Bell.   The several-pages-long
discussion of quantum incompleteness begins as follows:
\begin{quote}
\textcolor{OliveGreen}{``Physics is an attempt conceptually to grasp reality as it is thought
independently of its being observed.  In this sense one speaks of
`physical reality.'  In pre-quantum physics there was no doubt as to
how this was to be understood.  In Newton's theory reality was
determined by a material point in space and time; in Maxwell's theory,
by the field in space and time.  In quantum mechanics it is not so
easily seen.  If one asks:  does a $\psi$-function of the quantum
theory represent a real factual situation in the same sense in which
this is the case of a material system of points or of an
electromagnetic field, one hesitates to reply with a simple `yes' or
`no'; why?  What the $\psi$-function (at a definite time) asserts, is
this:  What is the probability for finding a definite physical
magnitude $q$ (or $p$) in a definitely given interval, if I measure it
at time $t$?  The probability is here to be viewed as an empirically
determinable, and therefore certainly as a `real' quantity which I may
determine if I create the same $\psi$-function very often and perform
a $q$-measurement each time.  But what about the single measured value
of $q$?  Did the respective individual system have this $q$-value even
before the measurement?''}
\end{quote}
The first sentence already makes perfectly clear that Einstein was not
using phrases like \textcolor{OliveGreen}{``real factual situation''} in the operationalist
sense that Wiseman's interpretation suggests.  And similarly,
Einstein's focus on \textcolor{OliveGreen}{``the single measured value''}
and 
\textcolor{OliveGreen}{``the individual
system''} make it clear that he is not merely interested in the
\textcolor{red}{``statistical''} type of effect that Wiseman
describes.  

It is worth continuing with Einstein's discussion.  Picking up where
the previous quote left off:
\begin{quote}
\textcolor{OliveGreen}{``To this question there is no definite answer within the framework of
the [existing] theory, since the measurement is a process which
implies a finite disturbance of the system from the outside; it would
therefore be thinkable that the system obtains a definite numerical
value for $q$ (or $p$)  the measured numerical value, only through the
measurement itself.  For the further discussion I shall assume two
physicists, A and B, who represent a different conception with reference
to the real situation as described by the $\psi$-function.}

\textcolor{OliveGreen}{``A.  The individual system (before the measurement) has a definite
value of $q$ (i.e., $p$) for all variables of the system, and more
specifically, \emph{that} value which is determined by a measurement
of this variable.  Proceeding from this conception, he will state:
The $\psi$-function is no exhaustive description of the real situation
of the system but an incomplete description; it expresses only what we
know on the basis of former measurements concerning the system.}

\textcolor{OliveGreen}{``B.  The individual system (before the measurement) has no definite
value of $q$ (i.e., $p$).  The value of the measurement only arises in
coorperation with the unique probability which is given to it in view
of the $\psi$-function only through the act of measurement itself.
Proceeding from this conception he will (or, at least, he may) state:
the $\psi$-function is an exhaustive description of the real situation
of the system.''}
\end{quote}
Einstein thus sets up a dilemma between two different views one might
take.  According to the ``A'' view, the distant system already
possesses definite, pre-determined values \textcolor{OliveGreen}{``for all variables''}.  We
may \emph{find out} the value of one of these variables by making an
appropriate sort of measurement on the entangled nearby system.  But
we do not influence or create those distant values.  Of course, the
existence of such pre-determined values requires us to say that the
$\psi$-function fails to provide a complete description of the real
physical state of the distant system.  

On the other hand, according to the ``B'' view, the $\psi$-function
can be claimed to provide a complete description of the real physical state of the
distant system because definite pre-determined values are no part of
that real physical state.  But then, as Einstein goes on to explain,
the quantum state $\psi_2$ of the distant system $S_2$ \textcolor{OliveGreen}{``depends upon
\emph{what kind} of measurement I undertake on $S_1$''}.  Continuing:
\begin{quote}
\textcolor{OliveGreen}{``Now it appears to me that one may speak of the real factual situation
of the partial system $S_2$.  Of this real factual situation, we know
to begin with, before the measurement of $S_1$, even less than we know
of a system described by the [original, pre-measurement]
$\psi$-function.  But on one supposition we should, in my opinion,
absolutely hold fast:  the real factual situation of the system $S_2$
is independent of what is done with the system $S_1$, which is
spatially separated from the former.  According to the type of
measurement which I make of $S_1$, I get, however, a very different
$\psi_2$ for the second partial system $(\phi_2, \phi_2^1, ...)$.
Now, however, the real situation of $S_2$ must be independent of what
happens to $S_1$.  For the same real situation of $S_2$ it is possible
therefore to find, according to one's choice, different types of
$\psi$-function.  (One can escape from this conclusion only by either
assuming that the measurement of $S_1$ ((telepathically)) changes the
real situation of $S_2$ or by denying independent real situations as
such to things which are spatially separated from each other.  Both
alternatives appear to me entirely unacceptable.)''  }
\end{quote}
What Einstein describes as
\textcolor{OliveGreen}{``unacceptable''} is unacceptable precisely in the sense of violating
the notion of locality that he has articulated previously (in the
sentence partially quoted by Bell).  So the upshot of Einstein's
discussion -- which Einstein goes on to state in the following
paragraph -- is that the ``B'' view described earlier is unacceptable
(i.e., non-local).  And that of course leaves only the ``A'' view,
which, remember, involves attributing definite pre-measurement values
\textcolor{OliveGreen}{``for all variables''} associated with the distant
system.  Einstein's conclusion is, in short, that the only way to
avoid an \textcolor{OliveGreen}{``unacceptable''} kind of nonlocality is to posit local
deterministic hidden variables.  

I have quoted and summarized this passage from Einstein's
``Autobiographical Notes'' at such length because it allows several crucial points to
be made about Bell's 1964 paper and Wiseman's interpretation thereof.
I have already  noted the complete implausibility of Wiseman's
operationalistic reading of Einstein.  Let us also now consider
Wiseman's charge that Bell \textcolor{red}{``made a mistake''} when he
(Bell) summarized Einstein \emph{et al.} as follows:
\begin{quote}
\textcolor{blue}{``Since we can predict in advance the result of
  measuring any chosen component of ${\bf{\sigma}}_2$ by previously
  measuring the same component of ${\bf{\sigma}}_1$, it follows that
  the result of any such measurement must actually be predetermined.''}
\end{quote}
Of course, Wiseman's claim that the argument sketched here is invalid
(\textcolor{red}{``a mistake''}) is based on Wiseman's interpretation
that Bell meant, by ``locality'', PI.  I have already explained why I
find that interpretation implausible.  

But how exactly does Wiseman's
accusation relate to Einstein and EPR?  As I pointed out above, it
seems (although it is admittedly not completely clear) that Wiseman
recognizes Einstein's conception of locality (and apparently also that
involved in the EPR argument) as broader than PI.
Wiseman also writes that \textcolor{red}{``EPR's premises ... are \emph{not} equivalent to
[Parameter Independence] and they \emph{do} justify the conclusion of
pre-determined outcomes.''}  So what exactly is the nature of the
``mistake'' that Wiseman is accusing Bell of having committed?

Is it that Bell was not attempting to \emph{rehearse} the earlier EPR
argument, but was instead attempting to \emph{replace} it with a new (and
invalid!) argument involving a narrower concept of ``locality''?   Or
is it that, although Bell was attempting to rehearse the earlier EPR
argument, he failed to capture it perfectly (in his two-sentence
recapitulation) and this supposed mistake somehow disqualifies that
aspect of his paper from consideration?  Or does Wiseman think that,
although an EPR-type argument from (something like Einstein's
generalized notion of) locality, to deterministic hidden variables,
\emph{can} be made rigorous, it was never made so until he, Wiseman,
made it so in his 2014 paper -- and \emph{that} is why
\textcolor{red}{``Bell's EPR paragraph forms no part of his 1964
  theorem''}?  I see no other available alternatives, yet none of
these are remotely reasonable as justifications for excluding, from
consideration, the EPR
part of Bell's two-part argument.  

Wiseman says that Bell's mistake has \textcolor{red}{``no impact on the main result
of his paper.''}  It is of course true that the EPR argument is
irrelevant to \textcolor{red}{``the main result''} if one arbitrary
stipulates \textcolor{red}{``the main result''} to be that the quantum
predictions are inconsistent with the joint assumptions of locality
and determinism.  But whether or not that is \textcolor{red}{``the
  main result''} is precisely what is fundamentally at issue in this debate, and
it should be clear that the EPR argument is quite crucial here.  If
the EPR argument (``\emph{from locality to} deterministic hidden
variables'') is valid, then \textcolor{red}{``the main result''} of
Bell's 1964 paper is that the quantum predictions are inconsistent
with the single unitary assumption of locality (meaning, of course,
the generalized notion of Einstein/EPR, not PI).  And the whole idea
that there are two distinct Bell's theorems falls apart.

Wiseman's suggestion that
orthodox quantum mechanics is some kind of
\textcolor{red}{``counter-example''} to Einstein's argument (which
Bell means to be summarizing) also underscores the implausibly
creative nature of Wiseman's interpretation.  
Einstein's entire several-page-long discussion (quoted above) 
is fundamentally \emph{about}
orthodox quantum mechanics and how it is, and isn't, possible to
understand that theory vis-a-vis locality and completeness.  
The idea that Einstein (or Bell) somehow made an argument for
locally pre-determined values, but without bothering to consider the
concrete example of orthodox quantum mechanics, is simply
ludicrous.   Einstein's whole argument is embedded in a discussion of
orthodox quantum mechanics from the very beginning.

Later in his paper, Wiseman considers (only to then dismiss it) the possibility
that Bell might have meant, by ``locality'', the condition articulated
by Einstein in the above-quoted passage where Einstein speaks of one
measurement \textcolor{OliveGreen}{``telepathically''} influencing the
other.  Wiseman writes:
\begin{quote}
\textcolor{red}{``Now although Bell seemed to indicate (twice) [sic] that
this [``no telepathy'' condition] was equivalent to his definition of locality, it
is different in that it requires not that Bob's \emph{result} $B$ be
independent of Alice's setting $a$, but rather that the `real
factual situation' of Bob's system be thus independent.''}  
\end{quote}
Indeed, as Wiseman proceeds to
acknowledge, Einstein's ``no telepathy'' notion of locality \textcolor{red}{``has the
same force as local causality"} and hence would support a valid
inference to deterministic hidden variables.\footnote{Strictly
  speaking this inference requires the
  additional assumption \textcolor{red}{``(which ... Einstein makes
    explicitly) that systems have real factual situations''}.  Recall
  that \textcolor{OliveGreen}{``denying independent real situations as
    such to things which are spatially separated from each other''} 
   was one of the two things that Einstein jointly
  described as \textcolor{OliveGreen}{``entirely unacceptable''}.  In
  my opinion, and probably that of Einstein, one must clearly accept
  that spatially-separated systems \emph{have} their own
  \textcolor{OliveGreen}{``real situations''} before one can even
  meaningfully ask whether locality is respected. (It is obvious, for
  example, that Einstein's formulation of locality -- the passage
  cited three times by Bell -- becomes incoherent if one does not
  already accept \textcolor{red}{``that systems have real factual
    situations.''})  The additional
  required assumption here would thus seem to be a logical
  precondition for discussing locality, rather than something one
  might coherently deny \emph{instead} of locality.  Note that this
  point is closely related to the important point that Bell would
  later express by insisting that the notion of locality must be
  formulated ``in terms of local beables.'' \cite{bell1976} }
But Wiseman dismisses this as irrelevant.  As we have already seen, Wiseman is
simply unwilling to believe Bell even when he (Bell) indicates repeatedly
that what he means by ``locality'' is what Einstein articulates in the
essay he cites.  And the existence
of a valid argument from Einstein's ``no telepathy'' version of
locality to deterministic hidden variables is
also supposedly irrelevant because, according to Wiseman, 
Einstein does not \textcolor{red}{``use it there [i.e., in his
``Autobiographical Notes''] to make the argument that Bell wants to
make, from predictability to determinism."}  But, as is plain from
the passages from Einstein's essay that I have quoted above, this is 
at best misleading.  Einstein's discussion in ``Autobiographical
Notes'' does indeed conclude that deterministic hidden-variables are
needed in order to avoid unacceptable non-locality.  One can dispute
the rigor with which that conclusion is argued for in that particular
discussion,\footnote{Einstein's argument in the ``Autobiographical Notes'' basically takes the following
  form:  either $A$ or $B$; not $B$; therefore $A$.  As I noted
  earlier, $A$ here includes
  the idea of deterministic hidden variables.  But as Matt Pusey
  correctly pointed out during the discussion of this paper in the
  IJQF ``John Bell Workshop 2014'',
  there are two slightly different sub-versions of $B$ -- one the
  denial of pre-determination, and one the more specific claim that
  quantum wave functions provide complete descriptions of physical
  states.  In the ``Autobiographical Notes'' Einstein gives a very
  clear argument, based on locality, against the second sub-version of
  $B$.  And of course the EPR paper provides an argument, again based
  on locality (although in the EPR text -- written by Podolsky -- the
  meaning and role of locality were not made particularly clear),
  against the first sub-version of $B$.  So it is probably correct to
  say that, taking the ``Autobiographical Notes'' alone, the argument
  for pre-determined values contains a kind of gap -- but also that,
  together, the ``Autobiographical Notes'' and the original EPR paper
  jointly contain \emph{precisely} the argument that
  \textcolor{red}{``Bell wants to make''}.}
but of course what Bell and all of us now have in mind
here is the EPR argument explicitly involving perfect correlations.

To sum up, Wiseman's interpretation requires us to believe things
about the views of both Einstein and Bell that are so completely at
odds with what is known generally about these thinkers -- and so
completely at odds with what they explicitly wrote in the specific
passages in question -- that I don't think it can be taken at all
seriously as capturing what Bell was actually doing in 1964.  Wiseman
writes, in a footnote, that \textcolor{red}{``there is no evidence to
  support the suggestion (Norsen, pers. comm.) that Bell began with a
  general notion of locality, along the lines of local causality, and
  only narrowed it to this definition after he had established
  determinism via his EPR paragraph.''}  I am truly at a loss to
understand how Wiseman could say this, since he himself has reviewed
the extensive and overwhelming evidence:  Einstein's formulation of locality, which Bell repeatedly cites,
  is precisely a \textcolor{red}{``general notion of locality, along
    the lines of local causality''}, quite distinct from PI, which
  allows a perfectly valid argument ``\emph{from locality to}
  deterministic hidden variables'', an argument which Einstein
  rehearses in the paragraphs immediately surrounding the sentence that
  Bell repeatedly cites and which Einstein \emph{et al.} gave in the
  EPR paper that Bell also cites. 

Wiseman, as far as I can tell, accepts all of this and yet still
somehow believes that the first paragraph of Bell's
\textcolor{blue}{``2 Formulation''} is merely \textcolor{red}{``a
  one-paragraph motivation for considering hidden variable
  theories.''}  I think it is clear that it is more than this.  It is
the first part of Bell's overall two-part argument.  It's just that
Bell is taking the first part as earlier work, as a
previously-established result that he need not rehearse in rigorous
detail, but may simply refer to and briefly summarize. 

This leaves, to my mind, only one question:  given that in 1964 Bell
presented his new result as the second part of a two-part argument for
the overall conclusion of non-locality (the first part of which was of
course the earlier EPR/Einstein argument), which portion, exactly, of this
two-part argument deserves to be called ``Bell's theorem''?  

Here I know from private communication that, when pressed in some of
the ways I've tried to lay out above, Wiseman retreats in the
direction of saying
that, while perhaps Bell may indeed have had the full two-part
argument in mind from the beginning, only the second part of it (the
part that was novel in 1964) deserves the epithet ``theorem''.   But
this strikes me as a terminological shell-game.  If a commentator wants
to reserve the word ``theorem'' for demonstrations meeting some
minimal threshold of rigor (and chooses to place the threshold
somewhere between the level found in Einstein's two cited papers and
what Bell did after the first paragraph of \textcolor{blue}{``2
  Formulation''} in his 1964 paper) I would have no objection, so long as
the commentator articulates clearly that ``Bell's theorem'', when combined with
the earlier ``EPR/Einstein non-theorem'' establishing the need for
deterministic hidden variables, leads to the overall conclusion that
the QM predictions are incompatible with locality... and that
\emph{this} is what Bell took himself to have established already in
1964.  I would even have no objection if such a commentator raised
questions about whether this incompatibility was really
\emph{established} in 1964, since (the commentator might plausibly
argue) genuinely \emph{establishing} such a conclusion requires that
all parts of the argument leading to it meet the commentator's threshold for
theoremhood.  What I do object to, however, is the gross
historical mischaracterization that is involved in
Wiseman's almost complete dismissal of the role of the EPR/Einstein argument (or
non-theorem or whatever one wants to call it) in Bell's 1964 paper.  At
the end of the day, and setting terminological games aside, Wiseman's
account of what Bell \emph{did} in 1964 is simply inaccurate in
that it fails to capture an \emph{essential} aspect of what Bell
actually established (and took himself to have
established).\footnote{As I pointed out in the IJQF ``John Bell
  Workshop 2014'' discussion of Wiseman's response to this paper, it
  is anachronistic to even obsess over what part, exactly, of what
  Bell did in 1964 ought to be included as part of ``Bell's Theorem.''
  There is of course no harm in saying that we are celebrating the
  50th anniversary of ``Bell's Theorem''.  But the truth is that we
  are celebrating the 50th anniversary of Bell's important 1964 paper.
  (The very phrase ``Bell's Theorem'' is a modern invention, which
  certainly played no role in Bell's thinking circa 1964.)}  

The long-standing disagreements about what Bell did, therefore,
cannot simply be understood as mere miscommunications, based on the
existence of two quite distinct ``Bell's theorems''.  The
disagreements are instead fundamentally based on the failure -- of 
Wiseman's ``operationalists'' and also apparently Wiseman himself -- 
to appreciate the foundational role of the
EPR/Einstein argument (``\emph{from locality to} deterministic hidden
variables'') in Bell's 1964 paper.  Wiseman's paper may perhaps be doing some
good in so far as his project involves making it more widely known
that Bell did \emph{eventually} establish a direct conflict between locality
(alone) and the quantum predictions.  But in so
far as his strategy involves telling the ``operationalists'' that they
were right all along, in how they understood Bell's 1964 paper,
Wiseman is distorting the historical record, muddying the waters, and
doing a great disservice to Bell on this 50th anniversary of his great
achievement.  

\section*{Acknowledgements}

Thanks to two anonymous referees, Daniel Tausk, Matt Pusey, and Howard
Wiseman for helpful comments about, and discussions based on, an
earlier draft of the paper.  And thanks to Roderich Tumulka for
helpful comments regarding Schilpp's translation of the crucial
passages from Einstein's ``Autobiographical Notes''.

\end{document}